\documentclass[prd,aps,nofootinbib,eqsecnum, superscriptaddress]{revtex4}

\usepackage{amsmath,amssymb,bm}
\usepackage[colorlinks]{hyperref}
\usepackage[all]{hypcap}
\usepackage{graphicx,psfrag,float,epsfig}
\usepackage{mathrsfs,wasysym}
\usepackage{epstopdf}
\usepackage{comment}
\usepackage{pbox}
\usepackage{array}
\usepackage{xspace}
\usepackage[usenames,dvipsnames]{color}

\newcommand{\nn}{\nonumber}

\def\ba{ {\bm a} }
\def\bL{ {\bm L} }
\def\bn{ {\bm n} }
\def\bp{ {\bm p} }
\def\bq{ {\bm q} }
\def\br{ {\bm r} }
\def\bS{ {\bm S} }
\def\bv{ {\bm v} }
\def\bx{ {\bm x} }

\def\bchi{ {\bm \chi} }
\def\bSigma{ {\bm \Sigma} }
\def\bxi{ {\bm \xi} }

\def\lbracket{ { \big \{ \!\! \big  \{ } }
\def\rbracket{ {\big \} \!\! \big \} } }

\def\addICTP{ICTP South American Institute for Fundamental Research,\\ Rua Dr. Bento Teobaldo Ferraz 271, 01140-070 S\~ao Paulo, SP Brazil}

\def\addIFT{Instituto de F\'isica Te\'orica, Universidade Estadual Paulista,\\ Rua Dr. Bento Teobaldo Ferraz 271, 01140-070 S\~ao Paulo, SP Brazil}

\def\addUPitt{Pittsburgh Particle Physics Astrophysics and Cosmology Center, Department of Physics and Astronomy, University of Pittsburgh, Pittsburgh, PA 15260, USA}

\def\addJPL{Jet Propulsion Laboratory, California Institute of Technology, Pasadena, CA 91109, USA}

\def\addCaltech{Theoretical Astrophysics, Walter Burke Institute for Theoretical Physics, California Institute of Technology, Pasadena, CA 91125, USA}

\begin{document}

\title{Radiation reaction for spinning bodies in effective field theory I:~Spin-orbit effects}

\author{Nat\'alia T.~Maia}
\affiliation{\addIFT}

\author{Chad R.~Galley}
\affiliation{\addJPL}
\affiliation{\addCaltech}

\author{Adam K.~Leibovich}
\affiliation{\addUPitt}

\author{Rafael A.~Porto}
\affiliation{\addIFT}
\affiliation{\addICTP}

\begin{abstract}
We compute the leading Post-Newtonian (PN) contributions at linear order in the spin to the radiation-reaction acceleration and spin evolution for binary systems, which enter at fourth PN order. The calculation is carried out, from first principles, using the effective field theory framework for spinning compact objects, in both the Newton-Wigner and covariant spin supplementary conditions. A non-trivial consistency check is performed on our results by showing that the energy loss induced by the resulting radiation-reaction force is equivalent to the total emitted power in the far zone, up to so-called ``Schott terms.''  We also find that, at this order, the radiation reaction has no net effect on the evolution of the spins. The spin-spin contributions to radiation reaction are reported in a companion~paper.
\end{abstract}

%\tableofcontents
\maketitle

%===================================
\section{Introduction}

%The dawn of gravitational wave astronomy with ground-based detectors \cite{Abbott:2016blz,2016htt,2016pea}, and future space-based antennas \cite{Sesana}, has motivated the computation of highly accurate theoretical models for the dynamics and gravitational wave emission from binary systems in general relativity. 
The potentially large amount of scientific information on strong gravitational fields that can be extracted from the observations of gravitational waves detected by ground-based detectors~\cite{Abbott:2016blz,2016htt,2016pea} and future spaced-based antennas motivates the computation of highly accurate theoretical models for the dynamics and gravitational wave emission from binary systems in general relativity. 
The gravitational dynamics and emitted power from compact binaries has been completed to seventh order in the relative speed $v$ (also called 3.5 post-Newtonian or 3.5PN order) for non-rotating bodies (see \cite{blanchet,Buoreview} for extensive reviews). In addition, the gravitational potential has been recently computed at 4PN order for non-spinning objects \cite{4pnjs,4pnjs2,4pndjs,4pndjs2,4pnbla,4pnbla2,4pndj} (see also \cite{lambzero,comp}).\vskip 4pt

The effective field theory (EFT) formalism, originally introduced in \cite{nrgr}, has readily reproduced many of these calculations, especially in the conservative sector for non-spinning bodies \cite{nrgr,nrgr2pn,nrgr3pn,nrgr4pn1,nltail,nrgr4pn2,tailfoffa}. On the other hand, when spin degrees of freedom are included~\cite{nrgrs}, the EFT framework has increased the knowledge of the gravitational dynamics and emitted power to 3PN order \cite{eih,3pnproc,comment, nrgrss,nrgrs2, nrgrso, srad, amps,Levi:2008nh,Perrodin:2010dy,Levi:2010zu}. The necessary ingredients were also computed in \cite{owen,buo1,buo2,damournloso,Schafer3pn,Schafer3pn2,schaferEFT,HergtEFT,bohennloss} using the  Arnowitt-Deser-Misner (ADM) and harmonic gauge formalisms. Higher order effects have been more recently studied in the conservative sector within the EFT approach in \cite{levinnlo1,levinnlo2,levinnlo3,equiv4pn} for the spin-orbit and spin-spin potentials at 3.5PN and 4PN, respectively. These results were also calculated in \cite{hartung, bohennloso,steinhoffnnlo1,luc12} with the ADM and harmonic methodologies,\footnote{Moreover, the next-to-leading order spin-orbit effects at 3.5PN order in the gravitational wave flux have been computed in \cite{luc13}, and the tail-induced contribution at 4PN in \cite{luc132}, using the harmonic approach.} except for the finite-size contributions, which are more efficiently handled by the EFT formalism \cite{nrgr, nrgrs,nrgrs2}. In addition, the effects due to absorption were studied within EFT in \cite{dis1,abspin}. For reviews of the EFT~framework~see~\cite{nrgrLH,rafric,iragrg,rafgrg,riccardocqg,review}.\vskip 4pt

The study of radiation reaction effects in gravity has also a celebrated history, notably the work of Burke and Thorne \cite{thorneBT1,thorneBT2}, who computed the leading order effects for non-spinning bodies at 2.5PN order. In the EFT approach, the incorporation of radiation reaction was developed in \cite{chadgsf,chadbr1} by implementing the classical limit of the ``in-in'' approach \cite{inin1, inin2}. The radiation reaction force to 3.5PN order, originally computed in \cite{Iyer1,luc93,Iyer2,luc96}, was subsequently re-derived with EFT methods in \cite{chadbr2} employing the formalism developed in \cite{chadprl, chadprl2} for nonconservative classical systems. (Higher order effects have been computed in \cite{Bala1} using a `refined balance' method.)
In the present work (part one of two) we incorporate radiation reaction effects for spinning bodies within an EFT framework from first principle, without resorting to balance equations. We compute the leading order spin-orbit radiation reaction acceleration and spin evolution at 4PN, which must be taken into account in order to construct fully accurate waveforms to this order. These effects were previously calculated in \cite{Will1} (and the radiation reaction Hamiltonian in \cite{RRADM}) using different methodologies and spin supplementarity conditions (SSCs).\footnote{See also \cite{G1,G2,WillBalance} for related work using balance equations.}  For completeness, here we carry out the calculation in both the Newton-Wigner and covariant SSCs. We also perform a non-trivial consistency check by showing the equivalence between the energy loss induced by the resulting radiation reaction acceleration and the total emitted power in the far zone which follows from the well-known multipole expansion, up to total time derivatives. (The latter are called ``Schott terms'' in electrodynamics \cite{schott}, which account for energy stored in near-zone fields. See also \cite{BDschott}.) We find that, in the spin-orbit sector, the radiation reaction has no net effect on the evolution of the spins. Our findings are compatible with the derivations in \cite{Will1}. Spin-spin back-reaction effects, which first enter at 4.5PN order, are studied in a companion~paper~\cite{paper2}.\vskip 4pt

As it is customary in the literature, we define the following useful quantities:
\begin{align}
	\bS \equiv {} & \bS_1 + \bS_2 , \\
	\bSigma \equiv {} & \frac{m}{m_2} \bS_2 - \frac{m}{m_1} \bS_1 , \\
	\bxi \equiv {} & \frac{m_2}{m_1} \bS_1 + \frac{m_1}{m_2} \bS_2 , \\
 	\bchi \equiv {} & \left(  2 + \frac{3m_2}{2m_1} \right)  \bS_1+\left(  2+\frac{3m_1}{2m_2} \right)  \bS_2 .
\end{align}
We also introduce 
\begin{align}
	\bL = \mu r \bn \times \bv ~, ~~ \tilde{\bL} \equiv \bL / \mu ,
\end{align}
with $m=m_1+m_2$, $\nu \equiv m_1 m_2 / m^2$, and $\mu = m\nu$. In addition, $\br \equiv \bx_1-\bx_2$ is the relative position, $\bv \equiv \bv_1-\bv_2$ and $\ba \equiv \ba_1 - \ba_2$ are the relative velocity and acceleration, respectively, and $\bn \equiv \br/r$. Throughout this paper we use $G_N=c=1$ units unless otherwise noted. We use the mostly minus signature convention for $\eta^{\alpha\beta}$ and Latin indices are contracted with the Euclidean metric.

%===================================
\section{Radiation-reaction in an EFT framework}

%-------------------------------------------------------------
\subsection{Non-spinning bodies}

Up to angular momentum and spin terms (see below), the worldline effective action describing the binary system in the radiation sector is given by \cite{andirad,andirad2}
\begin{align}
\label{seff0}
	S^{\rm rad}_{\rm eff} =  -\int dt \sqrt{\bar g_{00}} \bigg[M(t) -  \sum_{\ell=2} \bigg( \frac{1}{\ell!} I^L(t) \nabla_{L-2} E_{i_{\ell-1}i_\ell} +\frac{2\ell}{(2\ell+1)!}J^L(t) \nabla_{L-2} B_{i_{\ell-1}i_\ell}\bigg)\bigg] ,
\end{align}
with $L =(i_1\ldots i_\ell)$ being a multi-index. The quantity $M(t)$ is the (Bondi) mass associated with the binary while $I^L(t)$ and $J^L(t)$ are the mass- and current-type source multipole moments, respectively. The electric and magnetic components of the Weyl tensor are denoted by $E_{ij}$ and $B_{ij}$, respectively, which depend only on the metric in the radiation region, $\bar h_{\mu\nu}$. See \cite{review} for more details.

To incorporate the nonconservative effects of radiation reaction at the level of the action, we need to formally double the number of degrees of freedom~\cite{chadprl,chadprl2} so that $\bx_a \to ( \bx_{a (1)},\bx_{a(2)} )$ and $\bar{h}_{\mu\nu} \to ( \bar{h}_{\mu\nu}^{(1)}, \bar{h}_{\mu\nu}^{(2)} )$ in \eqref{seff0}. Here, $a=\{1,2\}$ labels the particles. Solving for both potential and radiation fields generates an effective action of the form
 \begin{align}
 \label{seff1}
	S_{\rm eff} [ \bx_{a \pm} ] = \int dt \, \big( L_{\rm eff} [\bx_{a (1)} ] - L_{\rm eff} [\bx_{a (2)} ] + R_{\rm eff} [ \bx_{a (1)}, \bx_{a (2)} ] \big),
\end{align}
where $L_{\rm eff} = \int dt (K - V)$ is the Lagrangian for the conservative sector and $R_{\rm eff}$ accounts for all nonconservative effects. It is convenient to translate the expressions into the $\pm$-variables, 
\begin{align}
	\bx_{a +} \equiv ( \bx_{a (1)} + \bx_{a (2) } ) /  2 ,\qquad \bx_{a -} \equiv \bx_{a (1) } - \bx_{a (2) }\,,\label{decomp}
\end{align}
such that the equations of motion follow from the variational principle:
\begin{equation}
\left[ \frac{\delta S_{\rm eff}[\bx_{a\pm}]}{\delta \bx_{a-} } \right]_{\rm PL} = 0\,. 
\end{equation}
The ``PL'' subscript indicates the ``physical limit'' is to be taken wherein all ``$-$'' variables vanish and the ``$+$'' variables are set to their physical values. In other words, in terms of the relative coordinates, $\br_-, \bv_- \to 0$, $\br_+ \to \br$, $\bv_+ \to \bv$,~etc. From here we obtain the (relative) acceleration due to radiation reaction, given by
\begin{align}
	\ba_{\rm RR}^{i} = \frac{1}{m\nu} \left[  \frac{\partial R_{\rm eff}(\br_\pm,\bv_\pm)}{\partial\br_{i-}(t)}-\frac{d}{dt}\left(  \frac{\partial R_{\rm eff}(\br_{\pm},\bv_\pm)}{\partial\bv_{i-}(t)}\right)  \right] _{\rm PL} .
	%_{\substack{\br_{-}^{i}=0\\\br_{+}^{i}=\br^{i}}}.
\label{eq:EL}
\end{align}

See~\cite{chadprl,chadprl2} for a more complete exposition of classical mechanics and field theories for generic, nonconservative systems.\vskip 4pt

At leading order in Newton's constant, the effective action is expressed in terms of Feynman diagrams as
\begin{align}
	\hspace{-5cm} iS_{\rm eff}[\bx_a^{(\pm)}] =   \sum_{\ell \ge 2}~~ \parbox{21mm}{\includegraphics[width=0.4\textwidth]{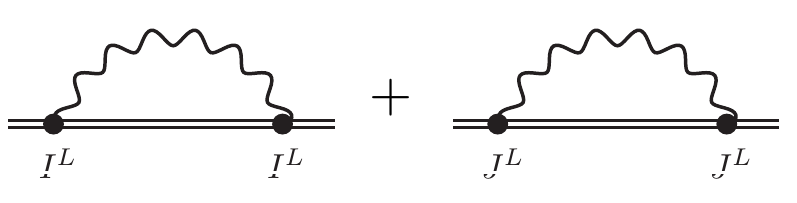}} \label{eq:Wtailall}
\end{align}
and yields the following nonconservative piece~\cite{nltail},
\begin{align}
\int R_{\rm eff}\, dt = {} & \sum_{\ell \ge 2}\, \frac{(-1)^{\ell+1}(\ell+2)}{(\ell-1)}  \int dt \left(\frac{2^\ell (\ell + 1)}{\ell (2 \ell + 1)!} I_{(-)}^L(t) I_{(+)}^{L \, (2\ell + 1)}(t) +  \frac{2^{\ell+3} \ell}{(2 \ell + 2)!}  J_{(-)}^L(t)  J_{(+)}^{L \, (2\ell + 1)}(t)\right) . \nn
\end{align}
The superscript $(n)$ in the multipole moments represents the number of time derivatives. 
For this paper, the lowest mass- and current-type multipole terms suffice to capture the contributions from the leading order spin effects to radiation reaction,
\begin{align}
\label{seff} 
	R_{\rm eff}[\br_\pm] = -\frac{1}{5} \, I_{-}^{ij}(t)I_{+}^{ij(5)}(t)-\frac{16}{45}\, J_{-}^{ij}(t) J_{+}^{ij(5)}(t).
\end{align}

For non-spinning bodies, we have at leading PN order the expressions for the mass quadrupole,
\begin{align}
	I_{(0)-}^{ij} (t) & \equiv I^{ij} (t; \bx_a^{(1)}) - I^{ij} (t; \bx_a^{(2)} )  =  \sum_a m_a \left(\bx_{a-}^{i} \bx_{a+}^{j} +  \bx_{a+}^{i} \bx_{a-}^{j} - \frac{2}{3}\, \delta^{ij} \bx_{a-} \! \cdot \! \bx_{a+}\right)  + {\cal O} (\bx_{a -}^3 ) ,  \\
	I_{(0)+}^{ij} (t) & \equiv \frac{1}{2} \Big( I^{ij} (t; \bx_a^{(1)}) + I^{ij} (t; \bx_a^{(2)} ) \Big) = \sum_a m_a \left( \bx_{a+}^i \bx_{a+}^j - \frac{1}{3} \delta^{ij} \bx_{a+}^2 \right) + {\cal O} (\bx_{a -}^2 )\,,
\end{align}
Using the first term of \eqref{seff} in \eqref{eq:EL}, we obtain the leading order radiation reaction piece of the full acceleration,
\begin{align}
	\left( \ba^{\rm RR}_{{\rm LO}a}\right)^i = -\frac{2}{5} \bx_a^j I^{ij(5)}_{(0)} .
\label{eq:LORR}
\end{align}
This result is the well-known Burke-Thorne acceleration~\cite{thorneBT1,thorneBT2} and was derived in the EFT approach in~\cite{chadbr1, chadbr2}.

%-------------------------------------------------------------
\subsection{Spinning bodies} 

%..........................................................................
\subsubsection{Basics}

The spin of an object is described by a $3$-vector. Therefore, a theory that is manifestly Lorentz invariant, which often represents spin by a tensor $S^{\mu\nu}$, must unavoidably introduce redundancies. To reduce the system to the required three degrees of freedom, a spin supplementary condition must be enforced. The two most popular SSCs are
\begin{align}
	S^{\mu\nu} p_\nu = {} & 0 ~, \qquad ({\rm Covariant}) \label{eq:covssc} \\
	S^{\mu 0} - S^{\mu j} \left( \frac{ \bp^j }{ p_0 + m } \right) = {} & 0 ~, \qquad ({\rm Newton-Wigner})  \label{eq:sscov}
\end{align}
where
\begin{align}
	p^{\alpha} = \frac{1}{\sqrt{u^2}} \left( m u^{\alpha} + \frac{1}{2m } R_{\beta\nu\rho\sigma}S^{\alpha\beta} S^{\rho\sigma} u^{\nu} + \cdots \right),  \label{eq:up}
\end{align}
with $u^\mu \equiv dx^\mu / d\sigma$, and $\sigma$ an affine parameter. At leading order we may write
\begin{align}
S^{i0} = \kappa \, S^{ij}\bv^j\,,
\label{eq:ssc}
\end{align}
with $\bv^j \equiv d\bx^j/dt$, and $\kappa = (1, 1/2 )$ in covariant and Newton-Wigner cases, respectively.

The SSCs are {\it second class} constraints, implying that their straightforward implementation in the effective action results in a modification of the symplectic structure, which introduces the so-called Dirac brackets, see e.g. \cite{HansonR}.
Nevertheless, we can retain the spin tensor until the end of the calculations, using instead a set of Lagrange multipliers. 
%In this framework, because the intrinsic angular variables do not participate in the dynamics, it is convenient to resort to an effective {\it Routhian}, in which the spin tensor is treated as a momentum variable. 
Because the spin degrees of freedom describe (angular) momentum, a partial Legendre transformation of the effective Lagrangian with respect to the spin tensors yields an effective {\it Routhian}.
In this framework, the equations of motion for the orbital dynamics and the spin dynamics are found by treating the Routhian as a Lagrangian and as a Hamiltonian, respectively~\cite{review}.
%The Routhian may be obtained following the same steps as in the non-spinning case, but the equations of motion for spin are derived as if we had a Hamiltonian instead \cite{review}.

The study of spin effects in the EFT framework was initiated in \cite{nrgrs,eih} in terms of an effective action approach, and subsequently elaborated in \cite{3pnproc,comment,nrgrss,nrgrs2,nrgrso,srad,amps} where the Routhian formalism was developed. (See \cite{yee} for earlier work in the context of motion in an external gravitaional field). For computational reasons (e.g.,~to~include spin-dependent finite size effects), the covariant SSC is more convenient, in which case the conservative dynamics of spinning bodies can be obtained from the Routhian\footnote{The $\omega_\mu{}^{ab}$ are the Ricci rotation coefficients and $R^\mu{}_{\alpha\beta\gamma}$ is the Riemann tensor.}
\begin{align}
\label{eq:Routh}
	{\cal R} =  - \left( m \sqrt{u^2} + \frac{1}{2} \omega_\mu{}^{ab} S_{ab} u^\mu + \frac{1}{2m }R_{\nu\alpha\rho\sigma}S^{\rho\sigma}u^{\nu} S^{\alpha\beta} u_\beta + \cdots\right)\,,
\end{align}
with $S^{ab}$ the spin tensor in a locally Minkowski frame described by the tetrad $e^a_\mu$, $S^{ab} \equiv S^{\mu\nu} e_\mu^a e_\nu^b$. The equations of motion follow from \begin{align}
	\frac{\delta }{\delta x^\mu} \int  {\cal R} ~ d\sigma = 0 ~, \qquad \frac{d
S^{ab}}{d\sigma} = \{ S^{ab}, {\cal R} \} \label{eq:eom},
\end{align}
where the orbital motion is derived with ${\cal R}$ playing the role of a Lagrangian (see the left equation above), while the spin dynamics is derived as if ${\cal R}$ were a Hamiltonian (see the right equation above).
Here, $\sigma$ may be chosen as the coordinate time, $t$, and the spin algebra is given by
\begin{align}
	\{ S^{ab},S^{cd} \} = \eta^{ac} S^{bd} +\eta^{bd}S^{ac}-\eta^{ad} S^{bc}-\eta^{bc}S^{ad}.\label{eq:algebra}
\end{align}
The last term(s) in \eqref{eq:Routh} ensures that the SSC is conserved during evolution.\vskip 4pt Because of the explicit breaking of Lorentz invariance in the Newton-Wigner SSC,  the form of the resulting Routhian is less compact. Moreover, the extra term to ensure the preservation of the SSC contributes already at leading order, unlike the one in \eqref{eq:Routh} which  enters at next-to-leading order (and only in the $\bS_a^2$ sector). However, when these terms (quadratic in the spin) are ignored, such as finite size effects, it turns out the Newton-Wigner SSC may be enforced prior to using \eqref{eq:eom}. The reason being that the resulting Dirac brackets turn into the canonical form, for instance for the spin\footnote{The minus sign is related to the mostly minus metric convention \cite{review}.}
\begin{align}
	\{\bS_{\rm (NW)}^i,\bS_{\rm (NW)}^j\} = -\epsilon^{ijk}\bS_{\rm (NW)}^k ,
\end{align}
unlike what occurs in the covariant case. This means that at linear order in the spin we may proceed {\it as usual}, after reducing the spin vector using \eqref{eq:ssc} with $\kappa = 1/2$. See \cite{review} and references therein for more details.\vskip 4pt

In what follows we will find it convenient to split the spin tensor into $3$-vector components,
\begin{align}
	\bS^i = \frac{1}{2} \epsilon^{ijk} S^{jk}\,,~~ \bS_{(0)}^i \equiv S^{i0} \,,
\end{align}
which are related via the SSC. For the covariant SSC, we will also have to consider the time variation of $\bS_{(0)}$, which may be obtained directly from the preservation of the SSC upon time evolution,
\begin{align}
\label{conssc}
	\dot \bS_{(0)a} \xrightarrow{\rm cov.\,\, SSC} \left(\ba_{a}\times \bS_a\right)^i +\cdots .
\end{align}

%..........................................................................
\subsubsection{Conservative dynamics}

In order to include all the spin-orbit effects at the desired PN order, we also need to account for the conservative part of the relative acceleration at linear order in the spin, which can be obtained in either SSC \cite{kidder,nrgrs,nrgrso},
\begin{align}
	\ba_{\rm cons}^{{\rm SO}({\rm NW})} = {} & \frac{1}{r^{3}}\left\{  \frac{3}{2}\bn\left[  \left(
\bn\times\bv\right)  \cdot\left(  7\bS+3\frac{\delta
m}{m}\bSigma\right)  \right]  -\bv\times\left(  7\bS 
+3\frac{\delta m}{m}\bSigma\right)  +\frac{3}{2}\dot{r}\left[
\bn\times\left(  7\bS+3\frac{\delta m}{m}\bSigma\right)
\right]  \right\} , \label{acnw} \\
	\ba_{\rm cons}^{{\rm SO}({\rm cov})} = {} & \frac{1}{r^{3}}\left\{  6\bn\left[  \left(
\bn\times\bv\right)  \cdot\left(  2\bS+\frac{\delta m}
{m}\bSigma\right)  \right]  -\bv\times\left(  7\bS
+3\frac{\delta m}{m}\bSigma\right) +3\dot{r}\left[  \bn
\times\left(  3\bS+\frac{\delta m}{m}\bSigma\right)  \right] 
\right\}  \label{acov}\,,
\end{align}
where $\delta m = m_1 - m_2$, and the spin variables on the right-hand side represent are in the Newton-Wigner and covariant SSC respectively.
These expressions enter at 1.5PN order in the conservative dynamics. For the spin we find the (conservative) evolution equations: \cite{kidder,nrgrs,nrgrso}
\begin{align}
	\dot\bS_{1\,\rm cons}^{\rm SO(NW)} = {} & \frac{1}{r^3} \left[2\left(1+\frac{3}{4}\frac{m_2}{m_1}\right) \bL\times \bS_{1\rm{(NW)}}\right] , \label{losnw} \\
	\dot \bS_{1\, \rm cons}^{\rm SO(cov)} = {} &  \frac{1}{r^3} \left[2\left(1+\frac{m_2}{m_1}\right) \bL\times \bS_{1 \rm{(cov)}} - m_2 (\bS_{1\rm{(cov)} }\times \br)\times \bv_1\right]\,, \label{loscov}
\end{align}
at linear order in the spins, for both the Newton-Wigner and covariant SSC respectively. Notice in both cases the spin evolution involves an extra factor of $v^2$, $\dot\bS \sim (v^3 / r) \bS$ (since $m/r \sim v^2$), which implies it may be taken as a constant vector at leading order. The expressions in \eqref{losnw} and \eqref{loscov} are related by the transformation~\cite{nrgrs,kidder}
\begin{equation}
\begin{aligned}
\label{covtonw}
	\bS_a \to {} & \left(1-\frac{\bv_a^2}{2}\right)\bS_a+ \frac{1}{2} (\bv_a \cdot \bS_a ) \bv_a ,  \\
	\bx_a \to {} & \bx_a + \frac{1}{2m_a} \bv_a \times \bS_a , 
\end{aligned}
\end{equation}
between Newton-Wigner and covariant SSCs.

%The expressions for the spin-orbit acceleration given in this section will be used in later sections to replace spin-independent terms that contain accelerations .

%We will use the expressions for the spin-orbit acceleration to incorporate all the radiation-reaction effects linear in the spin, in particular when we compute the time derivatives of spin-independent terms and replace them by the spin-dependent equations of motion. 

%..........................................................................
\subsubsection{Radiation reaction}

%In this paper we incorporate spin effects in radiation-reaction at linear order in the spins. We~will perform the computations in both the Newton-Wigner and covariant SSC. As we shall see, the latter case requires a few more subtleties. We~elaborate on the radiation-reaction formalism with spinning bodies in paper II, where we incorporate spin-spin and finite size effects.\vskip 4pt 

The incorporation of spin effects in radiation reaction requires doubling the number of degrees of freedom, not only for the positions and velocities but also for spin 
\begin{align} 
	S^{\mu\nu}_a \to \big( S^{\mu\nu}_{a (1)},S^{\mu\nu}_{a (2)} \big) .
\end{align}
Subsequently, it is useful to introduce the corresponding $\pm$-variables,
\begin{align}
\label{eq:SpinPlusMinus1}
	S^{\mu\nu}_{a +} \equiv ( S^{\mu\nu}_{a (1)} + S^{\mu\nu}_{ a (2) } ) /  2 , \qquad S^{\mu\nu}_{a -} \equiv S^{\mu\nu}_{a (1) } - S^{\mu\nu}_{a (2) } .
\end{align}
As before, after we solve for both for the potential and radiation fields in the far zone, we will find an expression similar to \eqref{seff1}, but this time for an effective Routhian with the following form,
 \begin{align}
 \label{Reff1}
	{\cal R}_{\rm eff} [ \bx_{a \pm} ,S^{\mu\nu}_{ a \pm} ] = {\cal R}^{\rm cons}_{\rm eff} [ \bx_{a(1)},S^{\mu\nu}_{a(1)}] - {\cal R}^{\rm cons}_{\rm eff} [ \bx_{a(2)},S^{\mu\nu}_{a(2)}] - {\cal R}^{\rm RR}_{\rm eff} [ \bx_{a \pm} ,S^{\mu\nu}_{ a \pm} ] .
\end{align}
The first terms account for corrections to the conservative sector whereas the last term, which cannot be written as a difference like the first two, incorporates radiation reaction effects due to spin. 
To the PN order we work in this paper, the dissipative term of the Routhian is given by
% entirely equivalent to the expression in \eqref{seff}, 
\begin{align}
\label{seffR} 
	{\cal R}^{\rm RR}_{\rm eff}[\br_\pm,S^{\mu\nu}_\pm] = -\frac{1}{5} \, I_{-}^{ij}(t)I_{+}^{ij(5)}(t)-\frac{16}{45}\, J_{-}^{ij}(t) J_{+}^{ij(5)}(t)\,,
\end{align}
including the spin-dependent contributions to the multipole moments. The acceleration describing radiation reaction is obtained from the effective Routhian in \eqref{seffR} by
\begin{align}
	\ba_{\rm RR}^{i} = \frac{1}{m\nu} \left[  \frac{\partial {\cal R}^{\rm RR}_{\rm eff}(\br_\pm,\bv_\pm,S^{\mu\nu}_\pm)}{\partial\br_{i-}(t)}-\frac{d}{dt}\left(  \frac{\partial {\cal R}^{\rm RR}_{\rm eff}(\br_{\pm},\bv_{\pm},S^{\mu\nu}_\pm)}{\partial\bv_{i-}(t)}\right)  \right]  _{\rm PL}  .
	%_{\substack{\br_{-}^{i}=0\\\br_{+}^{i}=\br^{i}}} . 
\label{eq:ER}
\end{align}

In order to fully describe the spin dynamics from the Routhian \eqref{seffR} we will need to extend the spin algebra to the $\pm$ variables. As discussed in \cite{chadprl}, to incorporate generic nonconservative effects the usual Poisson brackets must be generalized. In terms of the doubled phase space variables $(\bq_\pm, \bp_\pm)$, the new Poisson brackets are~\cite{chadprl}
\begin{align}
	\lbracket f,g \rbracket  \equiv \frac{\partial f}{\partial{\bq}_{+}} \cdot \frac{\partial g}{\partial{\bp}_{-}} 
		-\frac{\partial f}{\partial{\bp}_{-}} \cdot \frac{\partial g}{\partial{\bq}_{+}} 
		+ \frac{\partial f}{\partial{\bq}_{-}} \cdot \frac{\partial g}{\partial{\bp}_{+}} 
		- \frac{\partial f}{\partial{\bp}_{+}} \cdot \frac{\partial g}{\partial{\bq}_{-}} . 
\label{pb2}
\end{align}
To obtain the spin algebra we can proceed in two ways. We can either return to the original formulation  in \cite{nrgrs} (in terms of the angular velocity) and work out the steps to construct explicitly the spin brackets in the phase space, or we can simply use \eqref{pb2} to find the algebra for an angular momentum variable, as a generator of the Lorentz group, which must also be satisfied by a spin variable. In both cases we arrive at the following algebra:\footnote{As mentioned before, the minus sign is due to our convention for the Minkowski metric.}
\begin{equation}
\label{1s} 
\begin{aligned}
	\lbracket  \bS_{+}^{i},\bS_{+}^{j} \rbracket   = {} & -\frac{1}{4}\epsilon^{ijk}\bS_{-}^{k} ,\\
	\lbracket \bS_{-}^{i},\bS_{-}^{j} \rbracket   = {} & - \epsilon^{ijk}\bS_{-}^{k} , \\
	\lbracket \bS_{+}^{i},\bS_{-}^{j} \rbracket  = {} & - \epsilon^{ijk}\bS_{+}^{k} ,\\
	\lbracket \bS_{+}^{i},\bS_{(0)\pm}^j \rbracket  = {} & - \epsilon^{ijk}\bS_{(0)\mp}^k ,
\end{aligned}
\end{equation}
for the variables introduced in \eqref{eq:SpinPlusMinus1}. We will elaborate on the symplectic structure of these brackets elsewhere.\vskip 4pt
 
The equation of motion for spin will then be given by\footnote{Notice, since the brackets only affect the spin degrees of freedom, we can set to zero the minus variables associated with the orbital motion (e.g., $\br_- \to 0$) prior to computing \eqref{eq:SR}.}
\begin{align}
\label{eq:SR}
	\dot \bS_{\rm RR} = \lbracket \bS_{ +} ,{\cal R}^{\rm RR}_{\rm eff}[\br,\bS_{\pm},\bS_{(0) \pm}] \rbracket_{\rm PL} ,
\end{align}
where the physical limit ``PL" includes $\bS_{-}^{i} \to 0$ and $\bS_{+}^{i} \to \bS^{i}$.

%===================================
\section{Spin-orbit radiation reaction dynamics at 4PN order}

%-------------------------------------------------------------
\subsection{Source multipoles}\label{secm}

The multipole moments which are used to compute the spin-orbit radiation-reaction effects are given in \cite{srad,review}. In terms of the $\pm$-variables these are given by
\begin{align}
	I_{(0)-}^{ij} = {} & m\nu\left[  \br_{+}^{i}\br_{-}^{j}+\br_{-}^{i}\br_{+}^{j}\right]  _{\rm TF} , \\
	I_{(0)+}^{ij} ={} & m\nu\left[  \br_{+}^{i}\br_{+}^{j}\right]  _{\rm TF} , \\
	I_{S(0)-}^{ij} = {} & 2\nu \Bigg\{ m \Bigg[ \Bigg(  \frac{\bS_{(0)1+}^{i}}{m_1} - \frac{\bS_{(0)2+}^i}{m_2} \Bigg)  \br_{-}^{j} + \Bigg(  \frac{\bS_{(0)1-}^i}{m_1} - \frac{\bS_{(0)2-}^i}{m_2} \Bigg) \br_{+}^j \Bigg]+ \frac{1}{3} \epsilon^{ilk}\bxi_{-}^k\left(\bv^l_+\br^j_+ - 2 \br^l_+ \bv^j_+\right) \nn \\ 
		& +\frac{1}{3}\epsilon^{ilk}\bxi_+^{k}\left(\bv_{+}^{l}\br_{-}^{j} + \bv_{-}^{l}\br_{+}^{j}- 2 \br_{+}^{l}\bv_{-}^{j}-2\br_{-}^{l}\bv_{+}^{j}\right) \Bigg\}  _{\rm STF}, \\
	I_{S(0)+}^{ij} = {} & 2 \nu \left[m \left( \frac{\bS_{(0)1+}^i}{m_1} - \frac{\bS_{(0)2+}^i}{m_2}\right)  \br_{+}^{j} + \frac{1}{3}\epsilon^{ilk} \bxi^k_+\left(\bv_{+}^{l}\br_{+}^{j}-2\br_{+}^{l}\bv_{+}^{j}\right)\right]
_{\rm STF}, \\
	J_{(0)-}^{ij} = {} & -\nu\delta m\left[  \epsilon^{ikl}\left(  \br_{+}^{k}\bv_{+}^{l}\br_{-}^{j}+\br_{+}^{k}\bv_{-}^{l}\br_{+}^{j}+\br_{-}^{k}\bv_{+}^{l}\br_{+}^{j}\right)  \right]  _{\rm STF} , \\
	J_{(0)+}^{ij} = {} & -\nu\delta m\left[  \epsilon^{ikl}\br_{+}^{k} \bv_{+}^{l}\br_{+}^{j}\right]  _{\rm STF} , \\
	J_{S(0)-}^{ij} = {} & -\frac{3\nu}{2}\left[  \bSigma^{i}_+\br_{-}^j +  \bSigma^{i}_{-}\br_{+}^j \right]  _{\rm STF}\,, \label{Jmin} \\
	J_{S(0)+}^{ij} = {} & -\frac{3\nu}{2}\left[  \bSigma^{i}_{+}\br_{+}^{j}\right]  _{\rm STF} .
\end{align}
where ``(S)TF'' indicates the (symmetric) trace-free part of the quantity inside the brackets. We have ignored also terms involving products of minus variables (e.g., $\bS^i_{-} \br^j_{-}$), which do not contribute to the equations of motion or other physical quantities~\cite{chadprl}. Moreover, in the Newton-Wigner gauge we may apply the SSC prior to using \eqref{eq:ER} and \eqref{eq:SR}, in which case we have
\begin{align}
	I_{S(0)-}^{ij} \xrightarrow{\rm NW\,\, SSC} {} & \frac{\nu}{3} \left[\epsilon^{ikl}\bxi^{l}_+\left(
5\left(  \bv_{+}^{k}\br_{-}^{j}+\bv_{-}^{k}\br_{+}^{j}\right)  -4\left(  \br_{+}^{k}\bv_{-}^{j}+\br_{-}^{k}\bv_{+}^{j}\right) \right) + \epsilon^{ikl}\bxi^{l}_{-}\left(5\bv_{+}^{k}\br_{+}^{j}-4\bv_{+}^{k}\br^j_+\right) \right]_{\rm STF} , \label{Imin} \\
	I_{S(0)+}^{ij} \xrightarrow{\rm NW\,\, SSC} {} & \frac{\nu}{3} \left[  \epsilon^{ikl}\bxi^{l}_+\left(
5\bv_{+}^{k}\br_{+}^{j}-4\br_{+}^{k}\bv_{+}^{j}\right)  \right]  _{\rm STF} .
\end{align}

%-------------------------------------------------------------
\subsection{Acceleration}

In what follows we derive the accelerations in both the Newton-Wigner and covariant SSCs. In the former the spin tensor is reduced prior to applying \eqref{eq:ER} and \eqref{eq:SR} whereas in the latter the reduction is performed only after the equations of motion are obtained.

%..........................................................................
\subsubsection{Newton-Wigner SSC}

We split the computation into pieces, as in \cite{chadbr2}. The first term comes from the mass-quadrupole,
\begin{align}
	\ba_{{\rm RR}({\rm mq})}^{m}  = {} & -\frac{3}{5m}\left[  \epsilon^{ikl}\bv%
^{k}\bxi^{l}\delta^{jm}+\epsilon^{mil}\bxi^{l}\bv^{j}\right]  I_{(0)}^{ij(5)}-  \frac{2}{5}\left[  \br^{i}\delta^{jm}\right]I_{S(0)}^{ij(5)}  \\
	& -\frac{1}{15m}\left[ 5\epsilon^{mil}\bxi^{l}\br^{j}+4\epsilon^{ikl}\br^{k}\bxi^{l}\delta^{jm}\right]  I_{(0)}^{ij(6)} . \nn
\end{align}
After applying the equations of motion, we find
\begin{align}
	\ba_{{\rm RR}({\rm mq})} = {} & \frac{m\nu}{15r^{6}}\left\{  \left(  \tilde{\bL} \cdot \bxi \right) \left[  15\dot{r}\br\left(  42\frac{m}{r}-51v^{2}+119\dot{r}^{2}\right)  -2r\bv\left(  97\frac{m}{r}-81v^{2}
+405\dot{r}^{2}\right)  \right]  \right.  \\
		& -\left(  \br\times\bxi\right)  \left[  40\frac{m^{2}}{r}+261mv^{2}-15r\dot{r}^{2}\left(  41\frac{m}{r}-207v^{2}\right) - 333rv^{4}-3360r\dot{r}^{4}\right] \nn  \\
		&  \left.  -\dot{r}r^{2}\left(  \bv\times\bxi\right)  \left[596\frac{m}{r}-1233v^{2}+1905\dot{r}^{2}\right]  \right\} . \nn
\end{align}
The contribution from the current-quadrupole is
\begin{align}
	\ba_{{\rm RR}({\rm cq})}^{m} = {} & \frac{8}{15m}\left[ \bSigma^{i}\delta^{jm}\right]J_{(0)}^{ij(5)} + \frac{16\delta m}{45 m}\left[  \epsilon^{ikl}\br^{k}\bv^{l}\delta^{jm}+\epsilon^{imk}\left(  2\bv^{k}\br^{j} + \br^{k}\bv^{j}\right)  \right]J_{S(0)}^{ij(5)}  \\
		& + \frac{16\delta m }{45 m } \left[  \epsilon^{imk}\br^{k} \br^{j}\right]J_{S(0)}^{ij(6)} , \nn
\end{align}
yielding 
\begin{align}
	\ba_{{\rm RR}({\rm cq})} = {} & - \frac{4\nu\delta m}{15r^{6}}\left\{  \left(
\tilde\bL\cdot\bSigma\right)  \left[ 15\dot{r}\br \left(  12\frac{m}{r}-15v^{2}+35\dot{r}^{2}\right)  -4r\bv\left( 8\frac{m}{r}-9v^{2}+45\dot{r}^{2}\right)  \right]  \right.  \\
		&  -\left(  \br\times\bSigma\right)  \left[  22\frac{m^{2}}{r}-42mv^{2}+15r\dot{r}^{2}\left(  8\frac{m}{r}-9v^{2}\right)  +18rv^{4}+105r\dot{r}^{4}\right] \nn \\
		&  \left.  +15r^{2}\dot{r}\left(  \bv\times\bSigma\right) \left[  4\frac{m}{r}-3v^{2}+7\dot{r}^{2}\right]  \right\} \nn .
\end{align}
Finally, we have the ``reduced'' part from the leading order term in \eqref{seffR}. 
Keeping only the contribution from \eqref{acnw} in the time derivatives, we find
\begin{align}
	\ba_{{\rm RR}({\rm red})} = {} & \frac{2m\nu}{5r^{6}}\left\{  \left[  \left( \tilde\bL\cdot\bchi\right)  \right]  \left[  \dot{r}\br\left(  -32\frac{m}{r}-255v^{2}+455\dot{r}^{2}\right) + 4r\bv\left(  -7\frac{m}{r}+15v^{2}-60\dot{r}^{2}\right)  \right] \right.  \\
		& -3\left(  \br\times\bchi\right)  \left[  7mv^{2}+3r\dot{r}^{2}\left(  \frac{m}{r}+45v^{2}\right) -15rv^{4} -140r\dot{r}^{4}\right] \nn  \\
		& \left.  +5r^{2}\dot{r}\left(  \bv\times\bchi\right)  \left[ 4\frac{m}{r}+33v^{2}-45\dot{r}^{2}\right]  \right\}\,. \nn
\end{align}
The total 4PN radiation-reaction (relative) acceleration in the Newton-Wigner SSC is thus given by
\begin{align}
	 \left(\ba^{{\rm SO}({\rm NW})}_{\rm RR}\right)^{m}  = {} & -\frac{3}{5m}\left[  \epsilon^{ikl}\bv^{k}\bxi^{l}\delta^{jm}+\epsilon^{mil}\bxi^{l}\bv^{j}\right]  I_{(0)}^{ij(5)}-\frac{1}{15m}\left[  5\epsilon^{mil}\bxi^{l}\br^{j}+4\epsilon^{ikl}\br^{k}\bxi^{l}\delta^{jm}\right]  I_{(0)}^{ij(6)} \label{arrnw} \\
		&  -\frac{2}{5}\left[  \br^{i}\delta^{jm}\right]  I_{S(0)}^{ij(5)}-\frac{2}{5}\br^{j}\left[  I_{(0)}^{mj(5)}\right]  _{S}+\frac{8}{15m}J_{(0)}^{ij(5)}\left[ \bSigma^{i}\delta^{jm}\right]  \nn \\
		&  + \frac{16\delta m}{45m}\left[  \epsilon^{ikl}\br^{k}\bv^{l}\delta^{jm}+\epsilon^{imk}\left(  2\bv^{k}\br^{j}+\br^{k}\bv^{j}\right)  \right] J_{S(0)}^{ij(5)}+\frac{16\delta m}{45m}\left[  \epsilon^{imk}\br^{k}\br^{j}\right]  J_{S(0)}^{ij(6)} , \nn
\end{align}
which, after some algebra, becomes
\begin{align}
	\ba^{{\rm SO}({\rm NW})}_{\rm RR} = {} & \frac{2m\nu}{15r^{6}}\left\{  3\dot
{r}\br\left[  4\left( \tilde\bL\cdot\bS\right) \left(  14\frac{m}{r}-165v^{2}+315\dot{r}^{2}\right)  -9\left(  \tilde\bL\cdot\bxi\right)  \left(  7\frac{m}{r}+40v^{2}-70\dot{r}^{2}\right)  \right] \right.   \\
		&  -r\bv\left[  8\left( \tilde\bL\cdot\bS\right) \left(  29\frac{m}{r}-54v^{2}+225\dot{r}^{2}\right)  +3\left(  \tilde\bL\cdot\bxi\right)  \left(  53\frac{m}{r}-93v^{2}+375\dot{r}^{2}\right)  \right]  \nonumber \\
		&  -2\left(  \br\times\bS\right)  \left[  22\frac{m^{2}}{r}+21mv^{2}+3r\dot{r}^{2}\left(  49\frac{m}{r}+360v^{2}\right)  -117rv^{4} - 1155r\dot{r}^{4}\right] \nn \\
		&  +3\left(  \br\times\bxi\right)  \left[  8\frac{m^{2}}{r}-103mv^{2}+r\dot{r}^{2}\left(  169\frac{m}{r}-1215v^{2}\right)+135rv^{4}+1260r\dot{r}^{4}\right]  \nonumber \\
		&  \left.  +120r^{2}\dot{r}\left(  \bv\times\bS\right)  \left[ 9v^{2}-13\dot{r}^{2}\right]  -r^{2}\dot{r}\left( \bv\times\bxi\right)  \left[  88\frac{m}{r}-1269v^{2}+1755\dot
{r}^{2}\right]  \right\}\,\nn . 
\end{align}

%..........................................................................
\subsubsection{Covariant SSC}

Once again we split the radiation-reaction acceleration into separate terms. 
The only new expression relative to what we computed in the NW SSC, in terms of the source multipole moments, is given by the mass-quadrupole
\begin{align}
	\ba_{{\rm RR} ({\rm mq})}^{m}  = {} & -\frac{2}{5m}\left[  \left(  \frac{m}{m_{1}} \bS_{(0)1}^i-\frac{m}{m_{2}}\bS_{(0)2}^i\right)  \delta^{jm}+\epsilon^{mik}\bxi^{k}\bv^{j}+\epsilon^{ilk}\bv^{l}\bxi^{k}\delta^{jm}\right]  I_{(0)}^{ij(5)} \\
		& -\frac{2}{15m}\left[  \epsilon^{mik}\bxi^{k}\br^{j}+2\epsilon^{ilk}\bxi^{k}\br^{l}\delta^{jm}\right] I_{(0)}^{ij(6)}-\frac{2}{5}\left[  \br^{i}\delta^{jm}\right]  I_{S(0)}^{ij(5)} \nonumber
\end{align}
which becomes, after applying the equations of motion (and the covariant SSC prior to taking the time derivatives),\footnote{This is allowed since the SSC is conserved upon evolution.}
\begin{align}
	\ba_{{\rm RR} ({\rm mq})} = {} &-\frac{4m\nu}{15r^{6}}\left\{  \left(\tilde\bL\cdot\bxi\right)  \left[  15\dot{r}\br\left(  -6\frac{m}{r}-3v^{2}+7\dot{r}^{2}\right)  +2r\bv\left(  7\frac{m}{r}+9v^{2}-45\dot{r}^{2}\right)  \right]  \right. \\
		&  +\left(  \br\times\bxi\right)  \left[  8\frac{m^{2}}{r}+47mv^{2}+15r\dot{r}^{2}\left(  -7\frac{m}{r}+39v^{2}\right) - 63rv^{4}-630r\dot{r}^{4}\right]  \nonumber \\
		&  \left.  +\dot{r}r^{2}\left(  \bv\times\bxi\right)  \left[ 116\frac{m}{r}-243v^{2}+375\dot{r}^{2}\right]  \right\} . \nn
\end{align}
The current-quadrupole and reduced term remain formally the same as in the NW SSC. 
However, for the reduced part we input \eqref{acov} instead of \eqref{acnw} and obtain
\begin{align}
	\ba_{{\rm RR}({\rm red})}  = {} & -\frac{4m\nu}{15r^{6}}\left\{  \dot{r}\br \left[  \left( \tilde\bL\cdot\bS\right)  \left(  218\frac{m}{r}-675v^{2}+1435\dot{r}^{2}\right)  +\left( \tilde\bL \cdot\bchi\right)  \left(  -61\frac{m}{r}+720v^{2}-1400\dot{r}^{2}\right)  \right]  \right. \nonumber \\
		&  +3r\bv\left[  2\left( \tilde\bL\cdot \bS\right)  \left(  -21\frac{m}{r}+25v^{2}-115\dot{r}^{2}\right) +5\left( \tilde\bL\cdot\bchi\right)  \left(  7\frac{m}{r}-11v^{2}+47\dot{r}^{2}\right)  \right]  \nonumber \\
		&  +15\left(  \br\times\bS\right)  \left[  3mv^{2}+r\dot{r}^{2}\left(  -5\frac{m}{r}+27v^{2}\right)  -3rv^{4}-28r\dot{r}^{4}\right]  \nonumber \\
		&  +3\left(  \br\times\bchi\right)  \left[  3mv^{2}+r\dot{r}^{2}\left(  17\frac{m}{r}+135v^{2}\right)  -15rv^{4}-140r\dot{r}^{4}\right]  \nonumber \\
		&  \left.  +15r^{2}\dot{r}\left(  \bv\times\bS\right)  \left[ 4\frac{m}{r}-7v^{2}+11\dot{r}^{2}\right]  -15r^{2}\dot{r}\left( \bv\times\bchi\right)  \left[  4\frac{m}{r}+13v^{2}-17\dot{r}^{2}\right]  \right\}  .
\end{align}
Collecting all the pieces together, we find
\begin{align}
	\ba_{\rm RR}^{\rm SO(cov)}  = {} & \frac{2m\nu}{15r^{6}}\left\{  3\dot{r} \br\left[  4\left( \tilde\bL\cdot\bS\right)  \left( 14\frac{m}{r}-165v^{2}+315\dot{r}^{2}\right)  +\left( \tilde\bL \cdot\bxi\right)  \left(  \frac{m}{r}-540v^{2}+980\dot{r}^{2}\right) \right]  \right. \nonumber \\
		&  -r\bv\left[  8\left( \tilde\bL\cdot\bS\right) \left(  29\frac{m}{r}-54v^{2}+225\dot{r}^{2}\right)  +9\left( \tilde\bL \cdot\bxi\right)  \left(  31\frac{m}{r}-43v^{2}+175\dot{r}^{2}\right)  \right]  \nonumber\\
		&  -2\left(  \br\times\bS\right)  \left[ 22\frac{m^{2}}{r}+21mv^{2}+3r\dot{r}^{2}\left(  49\frac{m}{r}+360v^{2}\right)  -117rv^{4} - 1155r\dot{r}^{4}\right] \nonumber \\
		&  +\left(  \br\times\bxi\right)  \left[  28\frac{m^{2}}{r}-205mv^{2}+9r\dot{r}^{2}\left(  33\frac{m}{r}-295v^{2}\right) + 297rv^{4}+2730r\dot{r}^{4}\right]  \nonumber\\
		&  \left.  +120r^{2}\dot{r}\left(  \bv\times\bS\right)  \left[ 9v^{2}-13\dot{r}^{2}\right]  +r^{2}\dot{r}\left( \bv\times\bxi\right)  \left[  68\frac{m}{r}+981v^{2} - 1305\dot{r}^{2}\right]  \right\}  .
\end{align}

%-------------------------------------------------------------
\subsection{Spin evolution}

For the spin evolution we use \eqref{eq:SR} together with the Routhian in \eqref{seffR} and find
\begin{align}
	\dot {\bS}_{a\, \rm RR}^{k}  = -\frac{1}{5}\left[  I_{(0)+}^{ij(5)} \lbracket  \bS_{a +}^{k},I_{S(0)-}^{ij}\rbracket  +\frac{16}{9}J_{(0)+}^{ij(5)}\lbracket \bS_{a +}^{k},J_{S(0)-}^{ij} \rbracket  \right]_{\rm PL}.\label{eom:se}
\end{align}
To evaluate the right-hand side, we use the multipole moments from sec.~\ref{secm} and, depending on the SSC, the spin algebra in \eqref{1s}. 
As we show next, there is no radiation-reaction in the evolution for spin-orbit effects since the spin evolution equation can be written entirely as the time derivative of a redefined spin vector.

%..........................................................................
\subsubsection{Newton-Wigner SSC}

In this case we can enforce the SSC prior to applying the brackets. 
Using \eqref{Jmin} and \eqref{Imin}, we have (e.g., for particle~1)
\begin{align}
	\left(\dot{\bS}^{\rm SO(NW)}_{1 \, \rm RR}\right)^{k} = {} & - \frac{\nu}{15}\left[  I_{(0)}^{ij(5)}\left(
5\epsilon^{iql}{\bv}^{q}{\br}^{j}-4\epsilon^{iql}{\br}^{q}{\bv}^{j}\right)  \lbracket {\bS}_{ 1 +}^{k},{\bxi}_-^{l}\rbracket -8J_{(0)}^{ij(5)}{\br}^{j} \lbracket  {\bS}_{1 +}^{k},{\bSigma}_-^{i}\rbracket  \right]_{\rm PL} \label{spnweq} \\
		= {} & -\frac{\nu}{15}\left\{ \frac{m_2}{m_1} I_{(0)}^{ij(5)}\left[ -\bS_1^{i}\left(  5{\bv}^{k}{\br}^{j}-4{\br}^{k}{\bv}^{j}\right)  +\delta^{ik}\bS_1^{l}\left(  5{\bv}^{l}{\br}^{j}-4{\br}^{l}{\bv}^{j}\right)  \right] -8 \frac{m}{m_1}J_{(0)}^{ij(5)}{\br}^{j}\epsilon^{kiq}\bS^q_1 \right\} , \nn
\end{align}
for the radiation reaction evolution equation to linear order in the spins. Using the spin-independent equations of motion to reduce the derivatives on the multipoles we obtain
\begin{align}
	\dot {\bS}^{\rm SO(NW)}_{1\, \rm RR} = \frac{4m\nu\dot{r}}{15r^{4}} \left( {\bL}\times{\bS_1}\right) \left[ \left(  -22\frac{m}{r}+36v^{2}-60\dot {r}^{2}\right)  +\frac{m_2}{m_1}  \left( 16\frac{m}{r}-48v^{2}+75\dot{r}^{2}\right)  \right] .
\end{align}
It is easy to show the right-hand side is a total derivative that can be absorbed into a redefinition of the spin
\begin{align}
	\bS_1 \to \bS_1 - \frac{2m\nu}{15r^{3}} \left(  {\bL}\times{\bS_1}\right) \left[ \left(  3\frac{m}{r}-8v^{2} +24\dot{r}^{2}\right)  +\frac{m_2}{m_1} \left( \frac{m}{r}+12v^{2}-30\dot{r}^{2}\right)  \right] 
\end{align}
such that the new spin variable is insensitive to radiation reaction effects to 4PN order.

%..........................................................................
\subsubsection{Covariant SSC}

The spin evolution equation can also be obtained in the covariant SSC, using \eqref{eom:se} and the spin algebra in \eqref{1s}. Applying \eqref{conssc} after computing the brackets we obtain:
\begin{align}
\label{spcov}
	\left(\dot \bS^{\rm SO(cov)}_{1\, \rm RR}\right)^k = {} & -\frac{2\nu}{15}\left\{  I_{(0)}^{ij\left(5\right)  }\frac{m_{2}}{m_{1}}\left[  3\bS_1^kv^{i}{\br} ^{j} - 4\bv^k {\bS}_{1}^{i}{\br}^{j}+2{\bS}_{1}^{i}{\br}^{k}{\bv}^{j}+\delta^{ik}{\bS}_{1}^{l}\left(  {\bv}^{l}{\br}^{j}-2{\br}^{l}{\bv}^{j}\right)  \right]   -4\frac{m}{m_{1}}J_{(0)}^{ij\left(  5\right)  }\epsilon^{kil}{\br}^{j}{\bS}_{1}^{l}\right\} , 
\end{align}
leading to
\begin{align}
	\dot \bS^{\rm SO(cov)}_{1\, \rm RR} = {} & \frac{4m\nu}{15r^{4}} \bigg\{  \dot{r}\left({\bL}\times{\bS_1}\right) \left[  -22\frac{m}{r}+36v^{2}-60\dot{r}^{2}+\frac{m_{2}}{m_{1}}\left(  16\frac{m}{r}-48v^{2}+75\dot{r}^{2}\right)  \right]   \nn  \\
		& \qquad -\frac{m_{2}^{2}}{m} \bigg[  -2{\bS_1}\left(  6mv^{2}+r\dot{r}^{2}\left(  2\frac{m}{r}+99v^{2}\right)  -18rv^{4}-75r\dot{r}^{4}\right)  \nn  \\
		& \qquad +\dot{r}\left(  {\br}\left(  {\bS_1}\cdot{\bv}\right)  +{\bv}\left(  {\bS_1}\cdot{\br}\right)  \right)  \left(  2\frac{m}{r}+54v^{2}-75\dot{r}^{2}\right)   +6r{\bv}\left(  {\bS_1}\cdot{\bv}\right)  \left(  2\frac{m}{r}-6v^{2}+15\dot{r}^{2}\right)  \bigg]  \bigg\} .
\end{align}
The spin dynamics in the covariant SSC is not a total derivative due to the spin definition in this gauge.\footnote{This is not surprising since already at leading order the spin evolution equation for the covariant SSC does not conserve the norm of the vector, unlike the Newton-Wigner case. See \eqref{loscov} and \eqref{losnw}.}  However, it is easy to see the above result is equivalent to our derivation in the Newton-Wigner case. Recall that the transformation between the two spin variables is given by (see \eqref{covtonw})
\begin{align}
{\bS}_{1(\rm NW)}={\bS}_{1\rm (cov)} +\frac{1}{2}\Big(
{\bv}_{1} ({\bv}_{1}\cdot {\bS}_{1\rm (cov)})-{\bS}_{1\rm (cov)}{\bv}_{1}^2 \Big)+\cdots \,,
\end{align}
which implies
\begin{align}
	\dot {\bS}_{1\rm (NW)} = \dot {\bS}_{1\, \rm (cov)} - \frac{1}{2}\left(  \frac{m_{2}}{m}\right)^{2}\left[  2\, {\bS}_{1\, \rm (cov)}\left( {\ba} \cdot {\bv}\right)-\left(  {\bS}_{1\,\rm (cov)} \cdot {\bv}\right) {\ba}-{\bv} \left({\ba}\cdot {\bS}_{1\,\rm (cov)}\right)\right]+\cdots .
\end{align}
Hence, inputing the leading order dissipative part of the relative acceleration (see \eqref{eq:LORR}),
\begin{align}
	\left(\ba^{\rm RR}_{\rm LO}\right)^i = -\frac{2}{5} \br^j I^{ij(5)}_{(0)} ,
\end{align}
into the terms in the square brackets on the right hand side, and using \eqref{spcov}, we recover \eqref{spnweq} as expected.

%===================================
\section{Consistency test}

In this section we prove the equivalence between the power emitted at infinity via the multipole formula and the power induced by the radiation-reaction force, up to contributions that can be shown to be total time derivatives that average to zero for bound orbits.

%-------------------------------------------------------------
\subsection{Schott terms}

The radiated total power can be obtained from the effective action in \eqref{seff0} \cite{andirad,andirad2,review},
\begin{align}
	\frac{dE}{dt} =- \sum^{\infty}_{\ell=2} ~ \frac{(\ell+1)(\ell+2)}{\ell(\ell-1) \ell! (2\ell+1)!!}  \left( I^{L(\ell+1)}\right)^2 + \frac{4\,\ell(\ell+2)}{(\ell-1)(\ell+1)! (2\ell+1)!!} \left (J^{L(\ell+1)}\right)^2 ,\label{eq:power3}
\end{align}
which yields, to the order we work here,
\begin{align}
	\frac{dE}{dt} = - \frac{1}{5}\left[  I^{ij(3)}I^{ij(3)} + \frac{16}{9}  J^{ij(3)}J^{ij(3)} \right] +\cdots\,.\label{eq:powerLO}
\end{align}
The energy flux can be also obtained directly by computing the power induced by the radiation-reaction force,
\begin{align}
	{\cal P}_{\rm RR} \equiv m\nu\,\ba_{\rm RR}\cdot\bv . \label{powerso}
\end{align}
However, in one case the power is computed using the radiation field in the far zone, obtaining \eqref{eq:power3}, whereas the radiation-reaction force acts ``instantaneously'' on the dynamics of the bodies. The difference, nevertheless, is a local redefinition of the {\it conserved} energy (i.e., a total time derivative) that will not affect the radiated power in the far~region. These effects are often denoted as ``Schott'' terms, since they also appear in electrodynamics and, consequently, at leading order in the radiation-reaction force (e.g., see \cite{nltail}).

For non-rotating bodies the equivalence is almost straightforward. For instance, let us consider the leading order effective Lagrangian in \eqref{seff}. Then, according to the definition in \eqref{powerso} 
\begin{align}
	{\cal P}_{\rm RR} = -\frac{1}{5} \bv \cdot\left\{\left[  \frac{\partial}{\partial\br_{-}}-\frac{d}{dt} \frac{\partial}{\partial\bv_{-}} \right]\left(\, I_{-}^{ij}(t)I_{+}^{ij(5)}(t)+\frac{16}{9}\, J_{-}^{ij}(t) J_{+}^{ij(5)}(t)\right)\right\} _{\rm PL}. %_{\substack{\br_{-}=0\\\br_{+}=\br}}.
\end{align}
Crucially, the derivatives only act on the minus variables. Let us now take a time average of the above expression, 
\begin{align}
	\left\langle{\cal P}_{\rm RR}\right\rangle = {} & -\frac{1}{5}\left\langle\left( \left[\bv\cdot\frac{\partial}{\partial\br_{-}}+\ba\cdot\frac{\partial}{\partial\bv_{-}} \right]I_{-}^{ij}(t)\right) I_{+}^{ij(5)}(t)\right\rangle_{\rm PL} \\ % _{\substack{\br_{-} = 0 \\\br_{+}=\br}} \\ 
		&- \frac{16}{9}\,\left\langle \left(\left[\bv\cdot\frac{\partial}{\partial\br_{-}}+\ba\cdot\frac{\partial}{\partial\bv_{-}} \right]J_{-}^{ij}(t)\right) J_{+}^{ij(5)}(t)\right\rangle _{\rm PL} 
		%_{\substack{\br_{-}=0\\\br_{+}=\br}}
		\,,\nn
\end{align}
where we integrated by parts the time derivative in the second term in both of the square brackets. Noticing that
\begin{align}
	\left(\left[\bv\cdot\frac{\partial}{\partial\br_{-}}+\ba\cdot\frac{\partial}{\partial\bv_{-}} \right]I_{-}^{ij}(t)\right)_{\rm PL}%_{\substack{\br_{-}=0\\\br_{+}=\br}} 
	= I^{ij(1)}
\end{align}
we find
\begin{align}
	\left\langle{\cal P}_{\rm RR}\right\rangle = \left\langle\frac{dE}{dt}\right\rangle\,,
\end{align}
which implies
\begin{align}
\label{schott}
	{\cal P}_{\rm RR} = \frac{d\tilde E}{dt}\,,~~~\tilde E \equiv E-E_{\cal S}\,,
\end{align}
with $E_{\cal S}$ the Schott terms \cite{nltail}. The latter are given by 
\begin{align}
	E_{\cal S} = \frac{1}{5} \left(I^{ij(1)}I^{ij(4)} - I^{ij(2)}I^{ij(3)}\right) + \frac{16}{45} \left(J^{ij(1)}J^{ij(4)} - J^{ij(2)}J^{ij(3)}\right) + \cdots\,.
\end{align}

The equivalence, which can be proved to all orders, becomes a non-trivial consistency check when translated to the final expressions in terms of the variables of the problem.

%-------------------------------------------------------------
\subsection{Spinning bodies}

When spin is incorporated, the equivalence becomes a little less straightforward. The effective Lagrangian now depends on the spin tensor, $S^{ab}$, which is a new dynamical variable. However, at leading order, the spin vector simply plays the role of a constant source.\footnote{At lowest PN order the time variation of the spin vector is 1PN order higher than the naive power counting would suggest \cite{nrgrs,review}.} Hence, for the case of the Newton-Wigner SSC (where the spin tensor is reduced prior to obtaining the equations of motion) the above proof applies unaltered. This is not the case in the covariant gauge where the $S^{0i}$ components must be kept until the end of the calculation and the equation of motion for $S^{0i}$, which follows from the conservation of the SSC \eqref{conssc}, cannot be ignored. We explicitly demonstrate below how the consistency check applies in both cases.

%..........................................................................
\subsubsection{Newton-Wigner SSC}

The calculation of the spin-orbit radiation-reaction power in \eqref{powerso} is straightforward.
Using \eqref{arrnw} we obtain
\begin{align}
	{\cal P}^{{\rm SO(NW)}}_{\rm RR} = {} & \frac{4m\nu}{15r^{6}}\left\{  \left(  \bL\cdot\bS\right)  \left[  22\frac{m^{2}}{r}-95mv^{2}+3r\dot{r}^{2}\left(  77\frac{m}{r}-270v^{2}\right)  +99rv^{4}+735r\dot{r}^{4}\right] \right.  \\
		& \left.  -3\left(  \bL\cdot\bxi\right)  \left[  4\frac{m^{2}}{r}-25mv^{2}+4r\dot{r}^{2}\left(  29\frac{m}{r}-60v^{2}\right)+21rv^{4}+315r\dot{r}^{4}\right]  \right\}  , \nn
\end{align}
in the Newton-Wigner gauge. On the other hand, the radiated power in the far region reads
\begin{align}
	\left( \frac{dE}{dt}\right) _{{\rm SO (NW)}}=\frac{8m^{2}\nu}{15r^{6}}\left[  \left( \bL\cdot\bS\right)  \left(  12\frac{m}{r}+37v^{2}-27\dot{r}^{2}\right)  +\left(  \bL\cdot\bxi\right)  \left(  8\frac{m}{r}+19v^{2}-18\dot{r}^{2}\right)  \right] ,
\end{align}
which agrees with the literature \cite{kidder}. Comparing both expressions we find,
\begin{align}
	\left(  \frac{dE}{dt}\right)  _{{\rm SO(NW)}}- \mathcal{P}^{{\rm SO}}_{\rm RR} = {} & \frac{4m\nu}{15r^{6}}\left\{  \left(  \bL\cdot\bS\right)  \left[  2\frac{m^{2}}{r}+169mv^{2}+15r\dot{r}^{2}\left(  -19\frac{m}{r}+54v^{2}\right) - 99rv^{4}-735r\dot{r}^{4}\right]  \right.  \nonumber \\
		& \left.  +\left(  \bL\cdot\bxi\right)  \left[  28\frac{m^{2}}{r}-37mv^{2}+24r\dot{r}^{2}\left(  13\frac{m}{r}-30v^{2}\right) + 63rv^{4}+945r\dot{r}^{4}\right]  \right\} .
\end{align}
This can be shown to be a total time derivative, such that the expression in \eqref{schott} holds with a Schott term given by\footnote{The second derivative of the positions entering in the leading order multipole moment ($I^{ij}_{(0)}$ in the first term) must be replaced here only by the spin-orbit acceleration in~\eqref{acnw}.} 
\begin{align}
	E_{\cal S}^{\rm SO(NW)} = {} & \frac{1}{5}\left[I_{(0)}^{ij(1)} I_{(0)}^{ij(4)}-I_{(0)}^{ij(2)} I_{(0)}^{ij(3)}\right]_S +m\nu\left(  \bL\cdot\bS\right)  \left(  \frac{88}{15} \frac{m\dot{r}}{r^{5}}-\frac{72}{5}\frac{v^{2}\dot{r}}{r^{4}}+24\frac{\dot{r}^{3}}{r^{4}}\right)  \\
		& + m\nu\left(  \bL\cdot\bxi\right)  \left( -\frac{8}{3}\frac{m\dot{r}}{r^{5}}+\frac{129}{5}\frac{v^{2}\dot{r}}{r^{4}}-39\frac{\dot{r}^{3}}{r^{4}}\right) . \nonumber
\end{align}

%..........................................................................
\subsubsection{Covariant SSC}

The covariant case is a little more involved, since the binding energy now depends on $\bS_{(0)}$, which cannot be taken as a constant at leading order, unlike the spin $3$-vector. Therefore, the energy balance acquires an extra term compared to the NW calculation. At linear order in the spin we have
\begin{align}
\label{conE}
	\left\langle \left(\frac{dE}{dt}\right)_{\rm SO}\right\rangle = {} &\left\langle \left[  \frac{\partial E}{\partial \br^{i}}\bv^{i}+\frac{\partial E}{\partial \bv^{i}}\ba^{i}+\sum_a \frac{\partial E}{\partial \bS^i_{(0)a}}\dot \bS^i_{(0)a}\right]\right\rangle 
		= \left\langle m\nu\, \ba_{\rm SO(cov)}^{\rm RR}\cdot\bv +\sum_a \frac{\partial E}{\partial \bS^i_{(0)a}}\dot \bS^i_{(0)a}\right\rangle . 
\end{align}
Preserving the covariant SSC during evolution (see~\eqref{conssc}) implies
\begin{align}
	\dot \bS_{(0)a} \xrightarrow{\rm RR} \ba_{{\rm LO} a}^{\rm RR}\times \bS_a + \cdots ,
\end{align}
where we used only the {\it non-spinning} radiation-reaction part of the acceleration in \eqref{eq:LORR} since all other (conservative) terms cancel out in \eqref{conE}. On the other hand, the spin-orbit energy can be written as \cite{nrgrs,nrgrso,review}
\begin{align}
	E_{\rm SO} =-\sum_a \ba_{{\rm N} a}\cdot \bS_{(0) a} + \cdots ,
\end{align}
with $\ba_{{\rm N} a}$ the Newtonian acceleration for each body, which gives
\begin{align}
	\frac{\partial E_{\rm SO}}{\partial \bS^i_{(0)a} }=-\ba^i_{{\rm N}a} +\cdots .
\end{align}
From here we find
\begin{align}
	\frac{\partial E}{\partial \bS^i_{(0)a} } \dot\bS^i_{(0)a}    =-\sum_a \ba_{{\rm N}a}\cdot \left(\ba^{\rm RR}_{{\rm LO} a}\times \bS_a\right)   =-\frac{2\nu}{5}\int dt\epsilon^{ikl}\ba_{\rm N}^{k}\bxi^{l}\br^{j}I_{(0)}^{ij(5)} . \nonumber
\end{align}
Hence, the equivalence between far zone and radiation-reaction computation implies
\begin{align}
	\left(\frac{d\tilde E}{dt}\right)_{\rm SO(cov)} = \left[  m\nu\, \ba_{\rm RR}^{\rm SO(cov)}\cdot\bv-\frac
{2\nu}{5}\epsilon^{ikl}\ba_{\rm N}^{k}\bxi^{l}\br^{j}I_{(0)}^{ij(5)}\right] ,
\label{eqq}
\end{align}
with the introduction of Schott terms as in \eqref{schott}, which do not contribute to the time averaging. On the right hand side we have
\begin{align}
	\ba_{\rm RR}^{\rm SO(cov)}\cdot\bv = {} & \frac{4}{15r^{6}}\left\{  \left(  \bL\cdot\bS \right)  \left[  22\frac{m^{2}}{r}-95mv^{2}+3r\dot{r}^{2}\left(  77\frac{m}{r}-270v^{2}\right)  +99rv^{4}+735r\dot{r}^{4}\right]  \right. \nn \\
		& \left.  +\left(  \bL\cdot\bxi\right)  \left[  -14\frac{m^{2}}{r}-37mv^{2}-3r\dot{r}^{2}\left(  49\frac{m}{r}+90v^{2}\right) + 45rv^{4}+105r\dot{r}^{4}\right]  \right\}
\end{align}
and
\begin{align}
	\frac{2\nu}{5}\epsilon^{ikl}\ba^{k}\bxi^{l}\br^{j} I_{(0)}^{ij(5)}=-\frac{8m^{2}\nu}{5r^{6}}\left(  \bL \cdot\bxi\right)  \left(  2\frac{m}{r}-6v^{2}+15\dot{r}^{2}\right) .
\end{align}
On the other hand, the computation of the total radiated power using \eqref{eq:powerLO} yields,
\begin{align}
	\left(\frac{dE}{dt}\right)_{\rm SO(cov)} = \frac{8m^{2}\nu}{15r^{6}}\left[  \left(  \bL\cdot\bS\right)  \left(  12\frac{m}{r}+37v^{2}-27\dot{r}^{2}\right) + \left(\bL\cdot\bxi\right)  \left(  -4\frac{m}{r}+43v^{2}-51\dot{r}^{2}\right)  \right]  ,
\end{align}
also in agreement with the literature \cite{kidder}. It is then straightforward to show that the difference is a total time derivative, such that the expression in \eqref{eqq} holds with 
\begin{align}
	E_{\cal S}^{\rm SO(cov)} = {} & \frac{1}{5}\left[  I_{(0)}^{ij(1)} I_{(0)}^{ij(4)}-I_{(0)}^{ij(2)} I_{(0)}^{ij(3)}\right]_S + m\nu\left(  \bL\cdot\bS\right)  \left(  \frac{88}{15} \frac{m\dot{r}}{r^{5}}-\frac{72}{5}\frac{v^{2}\dot{r}}{r^{4}}+24\frac{\dot{r}^{3}}{r^{4} }\right)    \\
		& +m\nu\left(  \bL\cdot\bxi\right)  \left(  \frac{8}{5}\frac{m\dot{r}}{r^{5}}+36\frac{v^{2}\dot{r}}{r^{4}}-52\frac{\dot{r}^{3}}{r^{4}}\right)  ,\nonumber
\end{align}
for the Schott term.

%===================================
\section{Conclusions}

We incorporated radiation reaction effects due to spin in the dynamics of compact binary systems within the EFT formalism~\cite{review}. We extended the nonconservative approach developed in \cite{chadprl, chadprl2, chadbr2} to spinning bodies, and computed the spin-orbit contributions to the acceleration and spin evolution to 4PN order from first principles, both in the covariant and Newton-Wigner~SSCs. In order to test the consistency of our results, we explicitly showed that the induced power resulting from the radiation-reaction force is equivalent to the total radiated emission computed in the far zone, using the standard multipole expansion \cite{blanchet}. We find there is no net effect on the spin evolution from radiation reaction at this order, which is consistent with the findings in \cite{Will1}. The results presented here complete the knowledge of the binary's dynamics to 4PN order within an EFT framework~\cite{review}, which will play a key role in the forthcoming era of gravitational wave observations. Our work also paves the way for higher order calculations. We present the leading contributions from radiation reaction to the binary's dynamics due to spin-spin effects in a companion 
paper~\cite{paper2}.

%===================================
\section{Acknowledgements}

N.T.M. is supported by the Brazilian Ministry of Education (CAPES Foundation). 
C.R.G. is supported by NSF grant PHY-1404569 to Caltech.
A.K.L. is supported by the NSF grant PHY-1519175. 
R.A.P is supported by the Simons Foundation and S\~ao Paulo Research Foundation (FAPESP) Young Investigator Awards, grants 2014/25212-3 and 2014/10748-5. 
R.A.P. also thanks the theory group at DESY (Hamburg) for hospitality while this work was being completed.
Part of this research was performed at the Jet Propulsion Laboratory, California Institute of Technology, under a contract with the National Aeronautics and Space Administration.

\bibliography{RefSO}

\end{document}